\documentclass[usenatbib]{mn2e}
\usepackage{natbibmnfix,graphicx,times}
\usepackage{epstopdf}
\usepackage{amsmath,amssymb}

\newcommand{\apj}{ApJ}
\newcommand{\apjl}{ApJ}

\newcommand{\aj}{AJ}
\newcommand{\mnras}{MNRAS}
\newcommand{\physrep}{Physics Reports}
\newcommand{\prd}{Phys Rev D}

\newcommand{\iMpc}{\,\rm{Mpc^{-1}}}
\newcommand{\eV}{\, \rm{eV}}
\newcommand{\keV}{\, \rm{keV}}

\newcommand{\lya}{Ly$\alpha$ }

\newcommand{\lyb}{Ly$\beta$ }
\newcommand{\lybns}{Ly$\beta$}

\newcommand{\lyn}{Ly$n$ }

\newcommand{\deriv }{{\rm d}}
\newcommand{\ud }{{\rm d}}

\title[21 cm fluctuations from X-ray heating]{21 cm fluctuations from inhomogeneous X-ray heating before reionization}
\author[J.~R. Pritchard \& S.~R. Furlanetto]{Jonathan R. Pritchard$^1$\thanks{Email: jp@tapir.caltech.edu} and Steven R.  Furlanetto$^2$\thanks{Email: steven.furlanetto@yale.edu}\\
$^1$California Institute of Technology, Mail Code 130-33, Pasadena, CA 91125, USA\\
$^2$Yale Center for Astronomy and Astrophysics, Yale University, 260 Whitney Avenue, New Haven, CT 06520-8121, USA}

\begin{document}
\maketitle

\begin{abstract}
Many models of early structure formation predict a period of heating immediately preceding reionization, when X-rays raise the gas temperature above that of the cosmic microwave background.  These X-rays are often assumed to heat the intergalactic medium (IGM) uniformly, but in reality will heat the gas more strongly closer to the sources.  We develop a framework for calculating fluctuations in the 21 cm brightness temperature that originate from this spatial variation in the heating rate. High-redshift sources are highly clustered, leading to significant gas temperature fluctuations (with fractional variations $\sim 40\%$, peaking on $k\sim0.1\iMpc$ scales).  This induces a distinctive peak-trough structure in the angle-averaged 21 cm power spectrum, which may be accessible to the proposed Square Kilometre Array.  This signal reaches the $\sim10$ mK level, and is stronger than that induced by \lya flux fluctuations.  As well as probing the thermal evolution of the IGM before reionization, this 21 cm signal contains information about the spectra of the first X-ray sources.  Finally, we consider disentangling temperature, density, and \lya flux fluctuations as functions of redshift.
\end{abstract}


\section{Introduction} 
\label{sec:intro}
The formation of the first luminous objects ends the cosmic ``dark ages" and begins a period of heating and ionization of the intergalactic medium (IGM).  The global thermodynamic history of this epoch, which culminates in reionization, depends upon many poorly constrained processes such as star formation, radiative feedback, and the growth of HII regions \citep{bl2001}.  Currently, the best constraints on the ionization history come from observations of the Gunn-Peterson trough in quasar absorption lines \citep{gp1965} and in WMAP observations of the optical depth to recombination \citep{spergel2006}.  Current observations of the temperature evolution of the IGM are similarly limited.  At low redshift, observations of the \lya forest place constraints on the temperature of the IGM after reionization (\citealt{schaye2000,mcdonald2001}; \citealt*{zaldarriaga2001}; \citealt{theuns2002,hui2003}).  Unfortunately, photoionization during reionization causes a large temperature increase that essentially erases information about the preceding period.  At high redshift, it is assumed that the gas cools adiabatically after thermal decoupling from the CMB at $z\approx150$, when Compton scattering becomes inefficient \citep{peebles1993}.  The intermediate regime, where the first sources have ``switched on", is poorly constrained.  Once collapsed structures form many different heating mechanisms are possible, e.g., shock heating \citep{furlanetto2004}, resonant scattering of \lya photons \citep*{mmr1997,chen2004,chuzhoy2006heat,meiksin2006,rybicki2006,furlanetto2006heat}, and X-ray heating \citep*{ostriker1996,oh2001,venkatesan2001,ricotti2004}.  Determining the thermal history and identifying the important heating mechanisms requires new observations.

Future telescopes such as the {\em James Webb Space Telescope} hope to image high-redshift sources directly. However, seeing the sources is not the same as seeing the heating and ionization they cause in the IGM.  The most promising technique for probing the thermal history of the IGM before reionization is via observation of the 21 cm hyperfine transition of neutral hydrogen (\citealt{fob} and references therein).  This line may be seen in absorption against the CMB, when the spin  temperature $T_S$ is less than the CMB temperature $T_\gamma$, or in emission, when $T_S>T_\gamma$.  Three prototype low-frequency interferometers (LOFAR\footnote{See http://www.lofar.org/.}, MWA\footnote{See http://web.haystack.mit.edu/arrays/MWA/.}, and PAST\footnote{See \citet*{pen2005past}.}) are under construction and should be capable of observing the redshifted 21 cm signal from gas at redshifts $z\lesssim12$, with the proposed Square Kilometre Array\footnote{See \citet{carilli2004}.} (SKA) capable of probing even higher redshifts.  A great deal of theoretical work has now been done in calculating the 21 cm signal from fluctuations in density $\delta$ \citep{loeb_zald2004}, the \lya flux $J_\alpha$ \citep{bl2005detect,pritchard2006}, and the neutral fraction \citep*{zfh2004freq,fzh2004}.  Fluctuations in the 21 cm brightness temperature $T_b$ also occur because of fluctuations in the gas kinetic temperature $T_K$, but this has not yet been explored.

In this paper, we explore the effect of inhomogeneous X-ray heating by the first luminous sources on the 21 cm signal using analytic techniques.  We first build a model for the global thermal history of the IGM following \citet{furlanetto2006}.  In this model, we assume that a population of X-ray sources resulting from the remnants of the first stars is responsible for heating the IGM \citep{ostriker1996,oh2001,venkatesan2001,ricotti2004}.  X-ray heating is dominated by soft X-rays ($E\lesssim2 \keV$), as harder X-rays have a mean free path comparable with the Hubble scale.  These long mean free paths have often motivated the simplifying assumption that X-rays heat the IGM uniformly.  In fact, clustering of the sources and the $1/r^2$ decrease of flux with distance combine to produce significantly inhomogeneous heating.  We develop a formalism, based upon that of \citet{bl2005detect}, for calculating the temperature fluctuations that are sourced by these inhomogeneities.  We use this to explore features in the 21 cm power spectrum that constrain the evolution of $T_K$. This calculation also motivates a consideration of the possibility of using 21 cm measurements to constrain the X-ray emission spectrum of the first sources. 

Simulations of the early universe have yet to address the spectrum of temperature fluctuations in the period before reionization.  Previous analytic consideration of fluctuations in $T_K$ has focused on the period following recombination but before sources form \citep{bl2005infall, naoz2005}.  Temperature fluctuations induced by the first sources have not previously been considered in detail.

The 21 cm signal can be thought of as a tool for probing various radiation backgrounds.  Gas temperature fluctuations probe the X-ray background, neutral fraction fluctuations probe the ionizing UV background, and \lya fluctuations probe the non-ionizing UV background.  While the focus of this paper is X-ray heating of the IGM, in practice, the different sources of 21 cm fluctuation are not cleanly separated.  In order to properly establish context, we briefly re-examine the signal from fluctuations in the \lya flux, incorporating \lya production by X-ray excitation of HI \citep*{chen2006,chuzhoy2006}, and determine whether this contains extra useful information for constraining the spectral properties of the X-ray sources.  Finally, we explore the feasibility of separating information on the temperature and \lya flux fluctuations with the 21 cm signal. 
  
The layout of this paper is as follows.  We begin by setting out the physics of the 21 cm signal in \S \ref{sec:21cm}.  Calculating this requires a model for the global history of the IGM, which we outline in \S \ref{sec:global}.  Having established the mean history, in \S \ref{sec:fluctuate} we describe our framework for calculating fluctuations in $T_K$, $J_\alpha$, and the neutral fraction.  This is used to calculate the power spectrum for fluctuations in $T_K$ in \S\ref{sec:tkfluctuate}.  We then calculate the 21 cm signal in \S\ref{sec:tbfluctuate}, exploring the redshift evolution and dependence on the X-ray source spectrum and luminosity.  Finally, we discuss the possibility of observationally detecting and separating these signals in \S \ref{sec:observe} before concluding in \S \ref{sec:conclusion}.  Throughout this paper, we assume a cosmology with $\Omega_m=0.26$, $\Omega_\Lambda=0.74$, $\Omega_b=0.044$, $H=100h\,\rm{km\,s^{-1}\,Mpc^{-1}}$ (with $h=0.74$), $n_S=0.95$, and $\sigma_8=0.8$, consistent with the latest measurements \citep{spergel2006}, although we have increased $\sigma_8$ above the best-fit WMAP value to improve agreement with weak-lensing data.

\section{21 cm signal} 
\label{sec:21cm}

We begin by briefly summarising the physics of the 21 cm signal and refer the interested reader to \citet*{fob} for further information.  The 21 cm line of the hydrogen atom results from hyperfine splitting of the $1S$ ground state due to the interaction of the magnetic moments of the proton and the electron.  The HI spin temperature $T_S$ is defined via the number density of hydrogen atoms in the $1S$ singlet and triplet levels, $n_0$ and $n_1$ respectively, $n_1/n_0=(g_1/g_0)\exp(-T_\star/T_S)$, where $(g_1/g_0)=3$ is the ratio of the spin degeneracy factors of the two levels, and $T_\star\equiv hc/k\lambda_{21 \rm{cm}}=0.0628\,\rm{K}$.  The optical depth of this transition is small at all relevant redshifts, so the brightness temperature of the CMB is
\begin{multline}\label{brightnessT}
T_b=27  x_{\rm{HI}}(1+\delta_b)\\ \times\left(\frac{\Omega_bh^2}{0.023}\right)\left(\frac{0.15}{\Omega_mh^2}\frac{1+z}{10}\right)^{1/2}\left(\frac{T_S-T_\gamma}{T_S}\right)\,\rm{mK},
\end{multline}
Here $x_{\rm{HI}}$ is the neutral fraction of hydrogen and $\delta_b$ is the fractional overdensity in baryons.
The spin temperature is given by
\begin{equation}
T_S^{-1}=\frac{T_\gamma^{-1}+x_\alpha T_\alpha^{-1}+x_c T_K^{-1}}{1+x_\alpha+x_c},
\end{equation}
where $T_\alpha$ is the colour temperature of the \lya radiation field at the \lya frequency and is closely coupled to $T_K$ by recoil during repeated scattering.  The spin temperature becomes strongly coupled to the gas temperature when $x_{\rm{tot}}\equiv x_c+x_\alpha\gtrsim1$. 

The collisional coupling coefficient is given by
\begin{equation}
x_c=\frac{4T_\star}{3A_{10}T_\gamma}\left[\kappa^{HH}_{1-0}(T_k)n_H +\kappa^{eH}_{1-0}(T_k)n_e\right],
\end{equation}
where $A_{10}=2.85\times10^{-15}\,\rm{s}^{-1}$ is the spontaneous emission coefficient, $\kappa^{HH}_{1-0}$ is tabulated as a function of $T_k$ \citep{allison1969,zygelman2005} and $\kappa^{eH}_{1-0}$ is taken from \citet{furlanettobros}. For a more detailed analysis of the collisional coupling, see \citet{hirata2006col}. 

The Wouthysen-Field effect \citep{wouth1952,field1958} coupling is given by
\begin{equation}\label{xalpha}
x_\alpha=\frac{16\pi^2T_\star e^2 f_\alpha}{27A_{10}T_\gamma m_e c}S_\alpha J_\alpha,
\end{equation}
where $f_\alpha=0.4162$ is the oscillator strength of the \lya transition. $S_\alpha$ is a correction factor of order unity, which describes the detailed structure of the photon distribution in the neighbourhood of the \lya resonance \citep{chen2004,hirata2006lya,chuzhoy2006heat,furlanetto2006heat}.  We make use of the approximation for $S_\alpha$ outlined in \citet{furlanetto2006heat}.  For the models considered in this paper, \lya coupling dominates over collisional coupling.  

Fluctuations in the 21 cm signal may be expanded \citep{fob}
\begin{equation}\label{deltaTb}
\delta_{T_b}=\beta\delta+\beta_x\delta_x+\beta_\alpha\delta_\alpha+\beta_T\delta_T-\delta_{\partial v},
\end{equation}
where each $\delta_i$ describes the fractional variation in the quantity $i$: $\delta_\alpha$ for fluctuations in the Ly$\alpha$ coupling coefficient, $\delta_x$ for the neutral fraction, $\delta_T$ for $T_K$, and $\delta_{\partial v}$ for the line-of-sight peculiar velocity gradient. The expansion coefficients are given by
\begin{eqnarray}
\beta&=&1+\frac{x_c}{x_{\rm{tot}}(1+x_{\rm{tot}})},\\
\beta_x&=&1+\frac{x_c^{HH}-x_c^{eH}}{x_{\rm{tot}}(1+x_{\rm{tot}})},\nonumber\\
\beta_\alpha&=&\frac{x_\alpha}{x_{\rm{tot}}(1+x_{\rm{tot}})},\nonumber\\
\beta_T&=&\frac{T_\gamma}{T_K-T\gamma}\nonumber\\
&&+\frac{1}{x_{\rm{tot}}(1+x_{\rm{tot}})}\left(x_c^{eH}\frac{\deriv \log\kappa_{10}^{eH}}{\deriv \log T_K}+x_c^{HH}\frac{\deriv \log\kappa_{10}^{HH}}{\deriv \log T_K}\right)\nonumber.
\end{eqnarray}
In this, we assume that baryons trace the density field exactly so that $\delta_b=\delta$.
All of these quantities are positive, with the exception of $\beta_T$, whose sign is determined by $(T_K-T_\gamma)$.  The apparent divergence in $\beta_T$ when $T_K=T_\gamma$ is an artefact of expanding the fractional brightness temperature about a point where the mean brightness temperature $\bar{T}_b=0$.  The physical quantity $\bar{T}_b\beta_T$ is always well behaved.

Noting that in Fourier space $\delta_{\partial v}=-\mu^2\delta$ \citep{bharadwaj2004}, where $\mu$ is the angle between the line of sight and the wavevector $\mathbf{k}$ of the Fourier mode, we may use equation \eqref{deltaTb} to form the power spectrum \citep{bl2005sep}
\begin{equation}
P_{T_b}(k,\mu)=P_{\mu^0}(k)+\mu^2P_{\mu^2}(k)+\mu^4 P_{\mu^4}(k).
\end{equation}
In theory, high precision measurements of the 3D power spectrum will allow the separation of these terms by their angular dependence.  However, it is unclear whether the first generation of 21 cm experiments will be able to achieve the high signal-to-noise required for this separation \citep{mcquinn2005}.  Instead, they will measure the angle averaged quantity
\begin{equation}
\bar{P}_{T_b}(k)=P_{\mu^0}(k)+P_{\mu^2}(k)/3+P_{\mu^4}(k)/5.
\end{equation}
In presenting our results, we will concentrate on $P_{\mu^2}(k)$, which most cleanly separates out the different types of fluctuation, and $\bar{P}_{T_b}(k)$, which is easiest to observe.  We will typically plot the power per logarithmic interval $\Delta=[k^3 P(k)/2\pi^2]^{1/2}$.

\section{Global History} 
\label{sec:global}

\subsection{Outline}

We may express $T_b$ as a function of four variables $T_b=T_b(T_K,x_i,J_\alpha,n_H)$.  In calculating the 21 cm signal, we require a model for the global evolution of and fluctuations in these quantities.  We will follow the basic formalism of \citet{furlanetto2006}, but first let us consider the main events in likely chronological order. This determines redshift intervals where the signal is dominated by fluctuations in the different quantities.

$z\gtrsim200$: After recombination, Compton scattering maintains thermal coupling of the gas to the CMB, setting $T_K= T_\gamma$ so that we expect $\bar{T}_b=0$.

$40\lesssim z\lesssim200$: In this regime, adiabatic cooling means $T_K<T_\gamma$ and collisional coupling sets $T_S<T_\gamma$, leading to $\bar{T}_b<0$ and a possible absorption signal.  At this time, $T_b$ fluctuations are sourced by density fluctuations, potentially allowing cosmology to be probed \citep{loeb_zald2004,hirata2006col}.

$z_\star\lesssim z \lesssim 40$:  As the expansion continues, decreasing the gas density, collisional coupling becomes ineffective, absorption of CMB photons sets $T_S=T_\gamma$, and there is no detectable 21 cm signal.

$z_\alpha\lesssim z \lesssim z_\star$:  Once the first sources switch on at $z_\star$, they emit both \lya photons and X-rays.  In general, the emissivity required for \lya coupling is significantly less than that for heating $T_K$ above $T_\gamma$. Thus, in the simplest models, we expect the redshift $z_\alpha$, where \lya coupling saturates $x_\alpha\gg1$, to be greater than $z_h$, where $\bar{T}_K=T_\gamma$.  In this regime, $T_S\sim T_k<T_\gamma$ and there is an absorption signal.  Fluctuations are dominated by density fluctuations and variation in the \lya flux \citep{bl2005detect,pritchard2006,chen2006}.

$z_h \lesssim z \lesssim z_\alpha$:  After \lya coupling saturates, fluctuations in the \lya flux no longer affect the 21 cm signal.  By this point, heating becomes significant and gas temperature fluctuations source $T_b$ fluctuations. While $T_K$ remains below $T_\gamma$ we see a 21 cm signal in absorption, but as $T_K$ approaches $T_\gamma$ hotter regions may begin to be seen in emission.

$z_T \lesssim z \lesssim z_h$:  After the heating transition, $T_K>T_\gamma$ and we expect to see a 21 cm signal in emission.  The 21 cm brightness temperature is not yet saturated, which occurs at $z_T$, when $T_S\sim T_K\gg T_\gamma$.  By this time, the ionization fraction has likely risen above the percent level.  Brightness temperature fluctuations are sourced by a mixture of fluctuations in ionization, density and gas temperature.  

$z_r\lesssim z \lesssim z_T$:  Continued heating drives $T_K\gg T_\gamma$ at $z_T$ and temperature fluctuations become unimportant.  $T_S\sim T_K\gg T_\gamma$ and the dependence on $T_S$ may be neglected in equation \eqref{brightnessT}, which greatly simplifies analysis of the 21 cm power spectrum \citep{santos2006}.  By this point, the filling fraction of HII regions probably becomes significant and ionization fluctuations begin to dominate the 21 cm signal \citep{fzh2004}.

$z\lesssim z_r$:  After reionization, any remaining 21 cm signal originates from overdense regions of collapsed neutral hydrogen.

Most of these epochs are not sharply defined, so there should be considerable overlap between them.  This seems the most likely sequence of events, although there is considerable uncertainty in the ordering of $z_\alpha$ and $z_h$.  \citet{nasser2005} explores the possibility that $z_h>z_\alpha$, so that X-ray preheating allows collisional coupling to be important before the \lya flux becomes significant.  Simulations of the very first mini-quasar  \citep*{kuhlen2005,kuhlen2006} also probe this regime and show that the first luminous X-ray sources can have a great impact on their surrounding environment.  We note that these authors ignored \lya coupling, and that an X-ray background may generate significant \lya photons \citep{chen2006}, as we discuss in \S\ref{ssec:lyaflux}.

In this paper, we will concentrate on the period after $z_\star$, when luminous sources ``switch on", but before the IGM has been heated to temperatures $T_K\gg T_\gamma$ (our $z_T$).  In this regime, \lya coupling dominates and the 21 cm signal is seen in absorption at high $z$ but in emission at lower $z$.  We shall explore this transition in more detail below.  One of our key observables for 21 cm observations is the sign of $\beta_T$, which indicates whether $T_K>T_\gamma$ (provided that collisional coupling can be neglected).  

\subsection{Heating and ionization}

Having set the broad context, let us tighten our discussion with a concrete model for the evolution of the IGM;  in this we follow \citet{furlanetto2006}.  We will distinguish between the ionization fraction $x_i$, relating to the volume filled by the highly ionized HII regions that are located around clusters of sources, and the free electron fraction $x_e$ of the largely neutral gas outside these HII regions.  The former is important for determining when reionization occurs, while the latter governs X-ray heating in the bulk of the IGM.  We note that the volume filling fraction of the HII regions is well approximated by $x_i$, which we will use to calculate volume averaged quantities. 	 We further distinguish between $T_K$, the temperature of the IGM outside the HII regions, and the temperature of these photoionized regions $T_{\rm HII}\approx10^4{\rm \,K}$.  At high $z$, these regions are small and will not have a significant effect; while at low $z$, where reionization is well advanced, these HII regions will dominate and invalidate our formalism. 

We begin by writing down equations for the evolution of $T_K$, $x_i$, and $x_e$
\begin{eqnarray}
\frac{\deriv T_K}{\deriv t}&=&\frac{2T_K}{3n}\frac{\deriv n}{\deriv t}+\frac{2}{3 k_B}\sum_j \frac{\epsilon_j}{n},\label{thistory}\\
\frac{\deriv x_i}{\deriv t}&=&(1-x_e)\Lambda_i-\alpha_ACx_i^2n_H,\label{xihistory}\\
\frac{\deriv x_e}{\deriv t}&=&(1-x_e)\Lambda_e-\alpha_ACx_e^2n_H,\label{xehistory}
\end{eqnarray}
where $\epsilon_j$ is the heating rate per unit volume, and we sum over all possible sources of heating/cooling $j$.  We define $\Lambda_i$ to be the rate of production of ionizing photons per unit time per baryon applied to HII regions, $\Lambda_e$ is the equivalent quantity in the bulk of the IGM,  $\alpha_A=4.2\times10^{-13}\rm{cm^3\,s^{-1}}$ is the case-A recombination coefficient\footnote{Note that we use the case-A value, which amounts to assuming that ionizing photons are absorbed inside dense, neutral systems \citep{me2000}} at $T=10^4\,\rm{K}$, and $C\equiv\langle n_e^2\rangle/\langle n_e \rangle^2$ is the clumping factor.  We model the clumping factor using $C=2$; this value for $C$ reproduces the qualitative form of the histories in \citet{furlanetto2006} and ensures reionization at $z\gtrsim6$.  This approximation is appropriate only while $x_i$ is small, and will fail towards the end of reionization, when clumping becomes important in determining the effect of recombinations \citep*{me2000}.

In modelling the growth of HII regions, we take
\begin{equation}\label{lambdai}
\Lambda_i=\zeta(z)\frac{\deriv f_{\rm{coll}}}{\deriv t},
\end{equation}
where $f_{\rm{coll}}(z)$ is the fraction of gas inside collapsed objects at $z$ and the ionization efficiency parameter $\zeta$ is given by 
\begin{equation}
\zeta=A_{\rm{He}}f_\star f_{\rm{esc}}N_{\rm{ion}},
\end{equation}
with $N_{\rm{ion}}$ the number of ionizing photons per baryon produced in stars, $f_\star$ the fraction of baryons converted into stars, $f_{\rm{esc}}$ the fraction of ionizing photons that escape the host halo, and $A_{\rm{He}}$ a correction factor for the presence of Helium. This model for $x_i$ is motivated by a picture of HII regions expanding into neutral hydrogen \citep{bl2001}.  In calculating $f_{\rm{coll}}$, we use the \citet{ps1974mfn} mass function $\deriv n/\deriv m$ and determine a minimum mass $m_{\rm{min}}$ for collapse by requiring the virial temperature $T_{\rm{vir}}\ge10^4\,\rm{K}$, appropriate for cooling by atomic hydrogen. Decreasing this minimum mass, say to that of molecular cooling, will allow star formation to occur at earlier times shifting the features that we describe in redshift.  We note that $x_e\ll1$ at all redshifts under consideration, as once the free electron fraction reaches a few percent further X-ray energy is deposited primarily as heat, not further ionization.  

To integrate equation \eqref{thistory}, we must specify which heating mechanisms are important.
\citet{furlanetto2006} considers several heating mechanisms including shock heating \citep{furlanetto2004} and resonant scattering of \lya photons \citep{mmr1997,chen2004,chuzhoy2006heat,furlanetto2006heat}.  We shall neglect these contributions to heating of the IGM, focusing instead on the dominant mechanisms of Compton heating and X-ray heating.  While shock heating dominates the thermal balance at low $z$, during the epoch we are considering it, probably, heats the gas only slightly before X-ray heating dominates.

Compton heating serves to couple $T_K$ to $T_\gamma$ at redshifts $z\gtrsim 150$, but becomes ineffective below that redshift.  In our context, it serves to set the initial conditions before star formation begins.  The heating rate per particle for Compton heating is given by 
\begin{equation}\label{compton}
\frac{2}{3}\frac{\epsilon_{\rm{compton}}}{k_Bn}=\frac{x_e}{1+f_{\rm{He}}+x_e}\frac{T_\gamma-T_K}{t_\gamma}\frac{u_\gamma}{\bar{u}_\gamma}(1+z)^{4},
\end{equation}
where $f_{\rm{He}}$ is the helium fraction (by number), $u_\gamma$ is the energy density of the CMB, $\sigma_T=6.65\times10^{-25}\rm{cm^2}$ is the Thomson cross-section, and we define
\begin{equation}
t_\gamma^{-1}=\frac{8 \bar{u}_\gamma\sigma_T}{3m_e c}=8.55\times 10^{-13}\,\rm{yr}^{-1}.
\end{equation}

X-rays heat the gas primarily through photo-ionization of HI and HeI: this generates energetic photo-electrons, which dissipate their energy into heating, secondary ionizations, and atomic excitation.  With this in mind, we calculate the total rate of energy deposition per unit volume as
\begin{equation}
\epsilon_X=4\pi n_i\int \deriv \nu\, \sigma_{\nu,i}J_\nu (h\nu-h\nu_{\rm{th}}),
\end{equation}
where we sum over the species $i=$HI, HeI, and HeII, $n_i$ is the number density of species $i$, $h\nu_{\rm{th}}=E_{\rm{th}}$ is the threshold energy for ionization, $\sigma_{\nu,i}$ is the cross-section for photoionization, and $J_\nu$ is the number flux of photons of frequency $\nu$.  We may divide this energy into heating, ionization, and excitation by inserting the factor $f_i(\nu)$, defined as the fraction of energy converted into form $i$ at a specific frequency. The relevant division of the X-ray energy depends on $x_e$ and is calculated using the fitting formulae of \citet{SVS1985}.  The $f_i(\nu)$ are approximately independent of $\nu$ for $h\nu\gtrsim100\eV$, so that the ionization rate is related to the heating rate by a factor $f_{\rm{ion}}/(f_{\rm{heat}}E_{\rm{th}})$.  
The X-ray number flux is found from
\begin{eqnarray}\label{jxflux}
J_X(z)&=&\int_{\nu_{\rm{th}}}^{\infty}\deriv \nu\,J_X(\nu,z),\\
&=&\int_{\nu_{\rm{th}}}^{\infty}\deriv \nu\int_z^{z_{\star}} \deriv z'\,\frac{(1+z)^2}{4\pi}\frac{c}{H(z')}\hat{\epsilon}_X(\nu',z')e^{-\tau},\nonumber
\end{eqnarray}
where $\hat{\epsilon}_X(\nu,z)$ is the comoving photon emissivity for X-ray sources, and $\nu'$ is the emission frequency at $z'$ corresponding to an X-ray frequency $\nu$ at $z$
\begin{equation}\label{nup}
\nu'=\nu\frac{(1+z')}{(1+z)}.
\end{equation}
The optical depth is given by
\begin{multline}
\tau(\nu,z,z')=\int_z^{z'} \frac{\deriv l}{\ud z''}\ud z''\,[ n_{\rm{HI}}\sigma_{\rm{HI}}(\nu'')+n_{\rm{HeI}}\sigma_{\rm{HeI}}(\nu'')\\+n_{\rm{HeII}}\sigma_{\rm{HeII}}(\nu'')],
\end{multline}
where we calculate the cross-sections using the fits of \citet{verner1996}.  Care must be taken here, as the cross-sections have a strong frequency dependence and the X-ray frequency can redshift considerably between emission and absorption. In practice, the abundance of HeII is negligible and may be neglected.  

X-ray heating is often portrayed as uniform, as the X-ray photons possess long mean free paths.  The comoving mean free path of an X-ray with energy $E$ is \citep{fob}
\begin{equation}
\lambda_X\approx4.9 \bar{x}_{\rm{HI}}^{-1/3}\left(\frac{1+z}{15}\right)^{-2}\left(\frac{E}{300 \eV}\right)^3 {\rm\,Mpc} .
\end{equation}
Thus, the Universe will be optically thick, over a Hubble length, to all photons with energy below $E\sim2[(1+z)/15]^{1/2}\bar{x}_{\rm{HI}}^{1/3}\keV$.  The $E^{-3}$ dependence of the cross-section means that heating is dominated by soft X-rays, which do fluctuate on small scales.  In addition, though, there will be a uniform component to the heating from harder X-rays.  

We consider three possible sources of X-rays: starburst galaxies, supernova remnants (SNR), and miniquasars \citep{oh2001,glover2003,furlanetto2006}.  The incidences of starbursts and supernova remnants are likely to be tied to the global star formation rate \citep{glover2003}.  For simplicity, we will assume that miniquasars similarly track the star formation rate (SFR).  In reality, of course, their evolution could be considerably more complex \citep{madau2004}.  We characterise these sources by an emissivity per unit (comoving) volume per unit frequency
\begin{equation}\label{ehatX}
\hat{\epsilon}_X(z,\nu)=\hat{\epsilon}_X(\nu)\left(\frac{\rm{SFRD}}{\rm{M}_\odot \,\rm{yr}^{-1}\,Mpc^{-3}}\right),
\end{equation}
where SFRD is the star formation rate density, and the spectral distribution function is a power law with index $\alpha_S$
\begin{equation}
\hat{\epsilon}_X(\nu)=\frac{L_0}{h\nu_0} \left(\frac{\nu}{\nu_0}\right)^{-\alpha_S-1},
\end{equation}
and the pivot energy $h\nu_0=1\,\rm{keV}$.  We assume emission within the band 0.1 -- 30 keV, and set $L_0=3.4\times10^{40} f_X\,\rm{erg\,s^{-1}\,Mpc^{-3}}$, where $f_X$ is a highly uncertain constant factor.  This normalisation is chosen so that, with $f_X=1$, the total X-ray luminosity per unit SFR is consistent with that observed in starburst galaxies in the present epoch (see \citealt{furlanetto2006} for further details).  Extrapolating observations from the present day to high redshift is fraught with uncertainty, and we note that this normalisation is very uncertain.  The total X-ray luminosity at high redshift is constrained by observations of the present day soft X-ray back ground, which rules out complete reionization by X-rays, but allows considerable latitude for heating \citep*{dijkstra2004}.  Similarly, there is significant uncertainty in the spectra of these objects.  We choose $\alpha_S=1.5$ for starbursts, $\alpha_S=1.0$ for SNR, and $\alpha_S=0.5$ for miniquasars \citep{madau2004}.  These span the reasonable spectral dependence of possible X-ray sources.  

As in equation \eqref{lambdai}, we model the star formation rate as tracking the collapse of matter, so that we may write
the star formation rate per (comoving) unit volume
\begin{equation}
{\rm SFRD}=\bar{\rho}^0_b(z) f_{*}\frac{\deriv }{\deriv t}f_{\rm{coll}}(z).
\end{equation}
 where $\bar{\rho}^0_b$ is the cosmic mean baryon density today.  This formalism is appropriate for $z\gtrsim10$, as at later times star formation as a result of mergers becomes important.

\subsection{\lya flux}
\label{ssec:lyaflux}

Finally, we must describe the evolution of the \lya flux.  This is produced by stellar emission ($J_{\alpha,\star}$) and by X-ray excitation of HI ($J_{\alpha,X}$).  Photons emitted by stars, between \lya and the Lyman limit, will redshift until they enter a Lyman series resonance.  Subsequently, they may generate \lya photons, as discussed in \citet{pritchard2006} and \citet{hirata2006lya}.  The \lya flux from stars $J_{\alpha,\star}$ arises from a sum over the Ly$n$ levels, with the maximum $n$ that contributes $n_{\rm{max}}\approx23$ determined by the size of the HII region of a typical (isolated) galaxy (see \citealt{bl2005detect} for details).  The average \lya background is then
\begin{eqnarray}\label{jaflux}
J_{\alpha,\star}(z)&=&\sum_{n=2}^{n_{\rm{max}}}J^{(n)}_\alpha(z),\\
&=&\sum_{n=2}^{n_{\rm{max}}}f_{\rm{recycle}}(n)\nonumber\\
&&\times \int_z^{z_{\rm{max}}(n)} \deriv z'\,\frac{(1+z)^2}{4\pi}\frac{c}{H(z')}\hat{\epsilon}_\star(\nu'_n,z'),\nonumber
\end{eqnarray}
where $z_{\rm max}(n)$ is the maximum redshift from which emitted photons will redshift into the level $n$ Lyman resonance, $\nu'_n$ is the emission frequency at $z'$ corresponding to absorption by the level $n$ at $z$, $f_{\rm{recycle}}(n)$ is the probability of producing a \lya photon by cascade from level $n$, and $\hat{\epsilon}_\star(\nu,z)$ is the comoving photon emissivity for stellar sources.  We calculate $\hat{\epsilon}_\star(\nu,z)$ in the same way as for X-rays (eq. 21), and define $\hat{\epsilon}_\star(\nu)$ to be the spectral distribution function of the stellar sources.  We consider models with Pop. II and very massive Pop. III stars.  In each case, we take $\hat{\epsilon}_\star(\nu)$ to be a broken power law with one index describing emission between \lya and \lybns, and a second describing emission between \lyb and the Lyman limit (see \citealt{pritchard2006} for details).

Photoionization of HI or HeI by X-rays may also lead to the production of \lya photons.  In this case, some of the primary photo-electron's energy ends up in excitations of HI \citep{SVS1985}, which on relaxation may generate \lya photons \citep{mmr1997,chen2006,chuzhoy2006}.  This \lya flux $J_{\alpha,X}$ may be related to the X-ray heating rate as follows.  The rate at which X-ray energy is converted into \lya photons is given by
\begin{equation}
\epsilon_{X,\alpha}=\epsilon_{X,\rm{heat}} \frac{f_{\rm{ex}}}{f_{\rm{heat}}} p_\alpha,
\end{equation}
where $f_{\rm{ex}}$ and $f_{\rm{heat}}$ are the fraction of X-ray energy going into excitation and heating respectively, and $p_\alpha$ is the fraction of excitation energy that goes into \lya photons.  We then find the \lya flux by assuming that this injection rate is balanced by photons redshifting out of the \lya resonance, so
\begin{equation}
J_{\alpha,X}=\frac{c}{4\pi}\frac{\epsilon_{X,\alpha}}{h\nu_\alpha}\frac{1}{H\nu_\alpha}.
\end{equation}

\citet{SVS1985} calculated $f_{\rm{ex}}$ and $f_{\rm{heat}}$, but their Monte Carlo method, gives only a little insight into the value of $p_\alpha$.  Although excitations to the 2P level will always generate \lya photons, only some fraction of excitations to other levels will lead to \lya generating cascades.  The rest will end with two photon decay from the 2S level. \citet{SVS1985} considered a simplified atomic system, in which collisional excitations to $n\ge3$ levels were incorporated by multiplying the excitation cross-section to the $n=2$ level by a factor of 1.35 \citep{shull1979}.  Thus, we might expect of order $0.35/1.35\sim 0.26$ of collisional excitations to end at an $n\ge3$ level.  

We estimate $p_\alpha$ by calculating the probability that a secondary electron of energy $E_{\rm{sec}}$ will excite HI from the ground state to the level nL, using the collisional cross-sections\footnote{Taken from http://atom.murdoch.edu.au/CCC-WWW/index.html.} of \citet{bray2002}, and then applying the probability that the resulting cascade will produce a \lya photon, taken from \citet{pritchard2006} and \citet{hirata2006lya}.  The iterative procedure of \citet{pritchard2006} gives the probability of producing a \lya photon by cascade from the level nL as: (0, 1) for (2S, 2P), (1, 0, 1) for (3S, 3P, 3D), and (0.584, 0.261, 0.746, 1) for (4S, 4P, 4D, 4F).  

Summing over atomic levels $n\le4$, we obtain $p_\alpha=0.79$ for $E_{\rm{sec}}=30\eV$.  The contribution from $n>4$ levels is small as the collisional cross-sections drop off rapidly as $n$ increases.  The exact result depends upon the energy distribution of the secondary electrons, which in turn depends upon the spectrum of ionizing X-rays.  Our chosen value for $E_{\rm{sec}}$ corresponds to the mean electron energy (obtained using the distribution of \citealt{shull1979}) produced by X-rays of energy $1.7\keV$, which is the mean X-ray energy from a source with spectral index $\alpha=1.5$ over the band $0.1-30\keV$.  Calculating $p_\alpha$ exactly requires an update of the \citet{SVS1985} calculation, but, by considering different values for $E_{\rm{sec}}$, we conclude that it should differ from $p_\alpha=0.79$ by less than $10\%$. 

\subsection{Model histories}

\begin{figure}
\begin{center}
\includegraphics[scale=0.4]{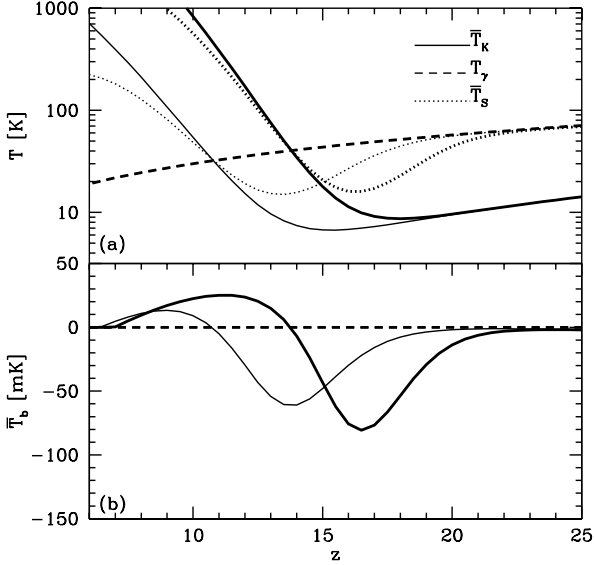}
\caption{Mean IGM thermal history for model A (thick curves) and B (thin curves). {\em (a): }$\bar{T}_K$ (solid curve), $T_\gamma$ (dashed curve), and $\bar{T}_S$ (dotted curve). {\em (b): }Volume averaged $\bar{T}_b$ (solid curve).  The zero line is indicated by a dashed horizontal line. Note that this is the thermal history outside of the ionized HII regions.}
\label{fig:global1}
\end{center}
\end{figure}
\begin{figure}
\begin{center}
\includegraphics[scale=0.4]{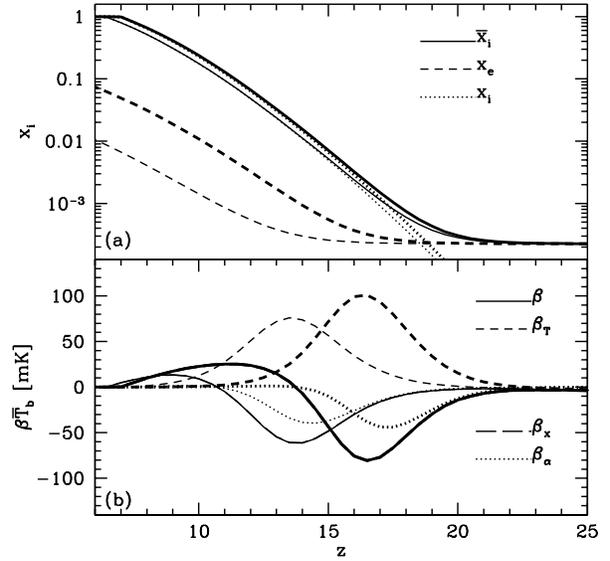}
\caption{Ionization histories for model A (thick curves) and B (thin curves). {\em (a): }$x_i$ (dotted curve), $x_e$ (dashed curve), and the volume averaged ionization fraction $\bar{x}_i=x_i+(1-x_i)x_e$ (solid curve). {\em (b): }The quantities $\beta_i \bar{T}_b$.  We plot $\beta$ (solid curve), $\beta_x$ (long dashed curve, overlapping with $\beta$), $\beta_T$ (short dashed curve), and $\beta_\alpha$ (dotted curve). }
\label{fig:global2}
\end{center}
\end{figure}
Having outlined the various elements of our global history, we will restrict ourselves to considering two models.  These will be A) Pop. II stars + starburst galaxies and B) Pop. III + starburst galaxies.  Of course, these are only two of an infinite set of possibilities, but they serve to illustrate the effect of different \lya and X-ray luminosities on the signal.  We use parameters corresponding to Pop. II ($f_{\rm{esc}}=0.1$, $f_\star=0.1$, $N_{\rm{ion}}=4000$) and very massive Pop. III  ($f_{\rm{esc}}=0.1$, $f_\star=0.01$, $N_{\rm{ion}}=30000$) stars \citep{furlanetto2006}, although we note that these values are highly uncertain.  We take $f_X=1$ in both models, to allow straightforward comparison between the two models.  The amplitude of the X-ray background is extremely uncertain, so that $f_X$ is almost unconstrained, and we defer discussion of its effects until \S\ref{ssec:xray}.

The global histories produced by these models are shown in Figure \ref{fig:global1} and Figure \ref{fig:global2}.  In Figure \ref{fig:global1}, we see the evolution of $\bar{T}_K$, $\bar{T}_S$, and $T_\gamma$.  Note that, while both models produce the same qualitative behaviour, the reduced star formation rate in Model B delays the onset of heating from $z\approx18$ to $z\approx15$.  We also see that the heating transition, where $\bar{T}_K=T_\gamma$, occurs at $z_h\approx14$ in Model A and $z_h\approx11$ in Model B.  We have assumed that the X-ray luminosity per unit star formation is the same for both populations, so this is an effect of $f_\star$ rather than the initial mass function.  In addition, Pop. III stars produce fewer \lya photons than Pop. II stars further slowing the onset of Lyman coupling.  

Figure \ref{fig:global1}b shows the distinctive $\bar{T}_b$ signature of absorption at $z>z_h$ followed by emission at $z<z_h$ in both models. The signal is significantly larger and more extended in Model A (See \citealt{furlanetto2006} for more detailed discussion of such histories).  The ionization history is outlined in Figure \ref{fig:global2}a and shows that $x_i$ evolves similarly in both models, as they have similar values for $\zeta$.  The electron fraction in the IGM $x_e$ is depressed in model B, where there is a smaller X-ray background.  Note that $x_e$ remains much smaller than $x_i$ once ionization begins. Both ionization histories produce an optical depth to the surface of last scattering $\tau_{ri}\approx0.07$, consistent with the WMAP third year observations of $\tau_{ri}=0.09\pm0.03$ \citep{spergel2006}, although slightly on the low side.   Our model for temperature fluctuations will be geared towards making predictions for the largely neutral IGM outside of the ionized HII regions surrounding clusters of UV sources.  Consequently, from Figure \ref{fig:global2}a, we expect our model to be valid for $z\gtrsim 12$, where $x_e\lesssim0.1$ and the filling fraction of the HII regions is small.

Figure \ref{fig:global2}b shows $\beta_i \bar{T}_b$, which is a measure of the sensitivity of the 21 cm signal to fluctuations in each fundamental quantity.  If the 21 cm signal were dominated by component $i$ and if the fluctuation had unit amplitude $\delta_i\approx1$, then $\beta_i \bar{T}_b$ gives the amplitude of the 21 cm signal.  Note that the curves for $\delta$ and $\delta_x$ are almost indistinguishable and track $\bar{T}_b$.   In contrast, the curves for $\delta_\alpha$ and $\delta_T$ show clear peaks -- representing windows where an existing signal might be seen.  We may identify $z_h$ as the point where $\bar{T}_b=0$ and all curves except that for $\beta_T$ go to zero.  At this point, the only fluctuations in $T_b$ arise from fluctuations in $T_K$.  In practice, this ``null" is more mathematical than physical, as inhomogeneities will blur the situation.  The redshift window for observing the 21 cm signal is clearly much narrower in model B, indicating that it will be much more confused than in model A.

\section{Formalism for temperature and ionization fluctuations} 
\label{sec:fluctuate}

Having specified our global history, we now turn to calculating the fractional fluctuations $\delta_\alpha$, $\delta_T$, and $\delta_x$.  Note that we will primarily be interested in the signal from the bulk of the IGM, working at redshifts where $x_i\lesssim0.1$, so that we will ignore the fluctuations induced by HII regions.  We begin by forming equations for the evolution of $\delta_T$ and $\delta_{e}$ (the fractional fluctuation in $x_e$) by perturbing equations \eqref{thistory} and \eqref{xehistory} (see also \citealt{bl2005infall,naoz2005}).  This gives 
\begin{equation}
\frac{\deriv \delta_T}{\deriv t}-\frac{2}{3}\frac{\deriv \delta}{\deriv t}=\sum_i\frac{2\bar{\Lambda}_{{\rm heat},i}}{3k_B\bar{T}_K} [\delta_{\Lambda_{{\rm heat},i}}-\delta_T],\label{evolve_deltaT}
\end{equation}
\begin{equation}
\frac{\deriv \delta_e}{\deriv t}=\frac{(1-\bar{x}_e)}{\bar{x}_e}\bar{\Lambda}_e[\delta_{\Lambda_e}-\delta_e]-\alpha_A C \bar{x}_e\bar{n}_H[\delta_e+\delta],\label{evolve_deltax}
\end{equation}
where an overbar denotes the mean value of that quantity, and $\Lambda=\epsilon/n$ is the ionization or heating rate per baryon.  We also need the fluctuation in the neutral fraction $\delta_x=-x_e/(1-x_e)\delta_{e}$ and in the \lya coupling coefficient $\delta_\alpha=\delta_{J_\alpha}$, neglecting the mild temperature dependence of $S_\alpha$ \citep{furlanetto2006heat}.

To obtain a closed set of equations, we must calculate the fluctuations in the heating and ionizing rates.  Perturbing equation \eqref{compton} we find that the contribution of Compton scattering to the right hand side of equation \eqref{evolve_deltaT} becomes \citep{naoz2005}
\begin{multline}\label{compton_perturb}
\frac{2\bar{\Lambda}_{{\rm heat,C}}}{3k_B\bar{T}_K}[\delta_{\Lambda_{{\rm heat,C}}}-\delta_T]=\frac{\bar{x}_e}{1+f_{\rm{He}}+\bar{x}_e}\frac{a^{-4}}{t_\gamma}   \\
\times \left[
4\left(\frac{\bar{T}_\gamma}{\bar{T}_K}-1\right)\delta_{T_\gamma}
+\frac{\bar{T}_\gamma}{\bar{T}_K}(\delta_{T_\gamma}-\delta_T)
\right],
\end{multline}
where $\delta_{T_\gamma}$ is the fractional fluctuation in the CMB temperature, and we have ignored the effect of ionization variations in the neutral fraction outside of the ionized bubbles, which are small.  Before recombination, tight coupling sets $T_K=T_\gamma$ and $\delta_T=\delta_{T_\gamma}$.  This coupling leaves a scale dependent imprint in the temperature fluctuations, which slowly decreases in time.  We will ignore this effect, as it is small ($\sim10\%$) below $z=20$ and once X-ray heating becomes effective any memory of these early temperature fluctuations is erased.  At low $z$, the amplitude of $\delta_{T_\gamma}$ becomes negligible, and equation \eqref{compton_perturb} simplifies.

Our main challenge then is to calculate the fluctuations in the X-ray heating.  We shall achieve this by paralleling the approach of \citet{bl2005detect} to calculating fluctuations in the \lya flux from a population of stellar sources.  We first outline their results (see also \citealt{pritchard2006}). Density perturbations at redshift $z'$ source fluctuations in $J_\alpha$ seen by a gas element at redshift $z$ via three effects.  First, the number of galaxies traces, but is biased with respect to, the underlying density field.  As a result an overdense region will contain a factor $[1+b(z')\delta]$ more sources, where $b(z')$ is the (mass-averaged) bias, and will emit more strongly.  Next, photon trajectories near an overdense region are modified by gravitational lensing, increasing the effective area by a factor $(1+2 \delta/3)$.  Finally, peculiar velocities associated with gas flowing into overdense regions establish an anisotropic redshift distortion, which modifies the width of the region contributing to a given observed frequency.  Given these three effects, we can write $\delta_{\alpha}=\delta_{J_\alpha}=W_\alpha(k)\delta$, where we compute the window function $W_{\alpha,\star}(k)$ for a gas element at $z$ by adding the coupling due to \lya flux from each of the \lyn resonances and integrating over radial shells \citep{bl2005detect} 
\begin{multline}\label{wk}
W_{\alpha,\star}(k)=\frac{1}{J_{\alpha,\star}}\sum_{n=2}^{n_{\rm{max}}}\int^{z_{\rm{max}}(n)}_z \deriv z' \frac{\deriv J^{(n)}_\alpha}{\deriv z'}\\
\times\frac{D(z')}{D(z)} \left\{[1+b(z')]j_0(kr)-\frac{2}{3}j_2(kr)\right\},
\end{multline}
where $D(z)$ is the linear growth function, $r=r(z,z')$ is the distance to the source, and the $j_l(x)$ are spherical Bessel functions of order $l$.  The first term in brackets accounts for galaxy bias while the second describes velocity  effects. The ratio $D(z')/D(z)$ accounts for the growth of perturbations between $z'$ and $z$. Each resonance contributes a  differential comoving \lya flux $\deriv J_\alpha^{(n)}/\deriv z'$, calculated from equation \eqref{jaflux}.  

\begin{figure}
\begin{center}
\includegraphics[scale=0.4]{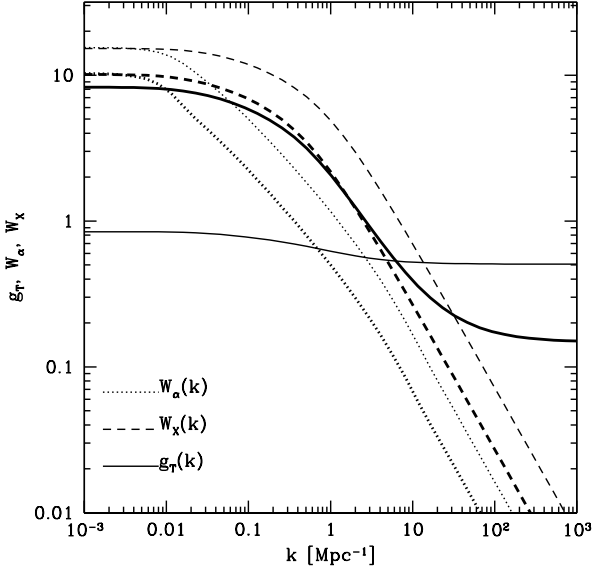}
\caption{$W_{\alpha,\star}(k)$ (dotted curves), $W_X(k)$ (dashed curves), and $g_T(k)$ (solid curves) at $z=20$ (thin curves) and $z=15$ (thick curves) for Model A.  }
\label{fig:wkcomp}
\end{center}
\end{figure}

We plot $W_{\alpha,\star}(k)$ in Figure \ref{fig:wkcomp}.  On large scales, $W_{\alpha,\star}(k)$ approaches the average bias of sources, while on small scales it dies away rapidly encoding the property that the \lya flux becomes more uniform.  In addition to the fluctuations in $J_{\alpha,\star}$, there will be fluctuations in $J_{\alpha,X}$.  We calculate these below, but note in passing that the effective value of $W_\alpha$ is the weighted average $W_\alpha=\sum_i W_{\alpha,i}(J_{\alpha,i}/J_\alpha)$ of the contribution from stars and X-rays. 

We now extend the formalism of \citet{bl2005detect} in an obvious way to calculate fluctuations in the X-ray heating rate. First, note that for X-rays $\delta_{\Lambda_{\rm{ion}}}=\delta_{\Lambda_{\rm{heat}}}=\delta_{\Lambda_\alpha}=\delta_{\Lambda_X}$, as the rate of heating, ionization, and production of \lya photons differ only by constant multiplicative factors (provided that we may neglect fluctuations in $x_e$, which are small).  In each case, fluctuations arise from variation in the X-ray flux.     We then write $\delta_{\Lambda_X}=W_X(k)\delta$ and obtain
\begin{multline}\label{wk_xray}
W_{X}(k)=\frac{1}{\bar{\Lambda}_{X}}\int_{E_{\rm{th}}}^{\infty} \deriv E\,\int^{z_{\star}}_z \deriv z' \frac{\deriv \Lambda_X(E)}{\deriv z'}\\ 
\times\frac{D(z')}{D(z)} \left\{[1+b(z')]j_0(kr)-\frac{2}{3}j_2(kr)\right\},
\end{multline}
where the contribution to the energy deposition rate by X-rays of energy $E$ emitted with energy $E'$ from between redshifts $z'$ and $z'+\deriv z'$ is given by
\begin{equation}
\frac{\deriv \Lambda_X(E)}{\deriv z'}=\frac{4\pi}{h}\sigma_{\nu}(E)\frac{\deriv J_X(E,z)}{\deriv z'}(E-E_{\rm{th}}),
\end{equation}
and $\bar{\Lambda}_X$ is obtained by performing the energy and redshift integrals.  Note that rather than having a sum over discrete levels, as in the \lya case, we must integrate over the X-ray energies.  The differential X-ray number flux is found from equation \eqref{jxflux}.

The window function $W_X(k)$ gives us a ``mask" to relate fluctuations to the density field; its scale dependence means that it is more than a simple bias.  The typical sphere of influence of the sources extends to several Mpc.  On scales smaller than this, the shape of $W_X(k)$ will be determined by the details of the X-ray source spectrum and the heating cross-section.  On larger scales, the details of the heated regions remain unresolved so that $W_X(k)$ will trace the density fluctuations.

A further word of explanation about this calculation is worthwhile. An X-ray is emitted with energy $E'$ at a redshift $z'$ and redshifts to an energy $E$ at redshift $z$, where it is absorbed.  To calculate $W_X$ we perform two integrals in order to capture the contribution of all X-rays produced by sources at redshifts $z'>z$.  The integral over $z'$ counts X-rays emitted at all redshifts $z'>z$ which redshift to an energy $E$ at $z$; the integral over $E$ then accounts for all the X-rays of different energies arriving at the gas element.  Together these integrals account for the full X-ray emission history and source distribution.  Many of these X-rays have travelled considerable distances before being absorbed.  The effect of the intervening gas is accounted for by the optical depth term in $J_X$.  Soft X-rays have a short mean free path and so are absorbed close to the source; hard X-rays will travel further, redshifting as they go, before being absorbed.  Correctly accounting for this redshifting when calculating the optical depth is vital as the absorption cross-section shows strong frequency dependence.  In our model, heating is dominated by soft X-rays, from nearby sources, although the contribution of harder X-rays from more distant sources can not be neglected.  

We compare the form of $W_X(k)$ and the stellar component of $W_\alpha(k)$ in Figure \ref{fig:wkcomp}. Including the X-ray contribution in $W_\alpha(k)$ drives that curve towards the $W_X(k)$ curve.  Note that $W_X$ shows significantly more power on smaller scales than $W_\alpha$, reflecting the greater non-uniformity in the X-ray heating; most heating comes from soft X-rays, which have mean free paths much smaller than the effective horizon of \lya photons.  Also, while $W_\alpha$ shows a subtle break in slope at $k\approx3\iMpc$, $W_X$ shows no obvious features indicative of preferred scales.  Both $W_X$ and $W_\alpha$ trace the bias on very large scales.

Returning now to the calculation of temperature fluctuations, to obtain solutions for equations \eqref{evolve_deltaT} and \eqref{evolve_deltax}, we let $\delta_T=g_T(k,z)\delta$, $\delta_e=g_e(k,z)\delta$, $\delta_\alpha=W_\alpha(k,z) \delta$, and $\delta_{\Lambda_X}=W_X(k,z)\delta$, following the approach of \citet{bharadwaj2004}.  Unlike \citet{bharadwaj2004}, we do not assume these quantities to be independent of scale, and so we must solve the resulting equations for each value of $k$.  Note that we do not include the scale dependence induced by coupling to the CMB \citep{naoz2005}. In the matter dominated limit, we have $\delta\propto(1+z)^{-1}$ and so obtain
\begin{equation}
\frac{\deriv g_T}{\deriv z}=\left(\frac{g_T-2/3}{1+z}\right)-Q_X(z)[W_{X}(k)-g_T]-Q_C(z)g_T,\label{evolve_gT2}
\end{equation}
\begin{equation}
\frac{\deriv g_e}{\deriv z}=\left(\frac{g_e}{1+z}\right)-Q_I(z)[W_{X}(k)-g_e]+Q_R(z)[1+g_e],\label{evolve_ge2}
\end{equation}
where we define
\begin{equation}
Q_I(z)\equiv\frac{(1-\bar{x}_e)}{\bar{x}_e}\frac{\bar{\Lambda}_{{\rm ion},X}}{(1+z)H(z)},
\end{equation}
\begin{equation}
Q_R(z)\equiv\frac{\alpha_A C \bar{x}_e\bar{n}_H}{(1+z)H(z)},
\end{equation}
\begin{equation}
Q_C(z)\equiv\frac{\bar{x}_e}{1+f_{{\rm He}}+\bar{x}_e}\frac{(1+z)^3}{t_\gamma H(z)}  \frac{T_\gamma}{\bar{T}_K},
\end{equation}
and
\begin{equation}
Q_X(z)\equiv\frac{2\bar{\Lambda}_{{\rm heat},X}}{3k_B\bar{T}_K(1+z)H(z)}.
\end{equation}
These are defined so that $Q_R$ and $Q_I$ give the fractional change in $x_e$ per Hubble time as a result of recombination and ionization respectively.  Similarly, $Q_C$ and $Q_X$ give the fractional change in $\bar{T}_K$ per Hubble time as a result of Compton and X-ray heating.  Immediately after recombination $Q_C$ is large, but it becomes negligible once Compton heating becomes ineffective at $z\sim150$.  The $Q_R$ term becomes important only towards the end of reionization, when recombinations in clumpy regions slows the expansion of HII regions.  Only the $Q_X$ and $Q_I$ terms are relevant immediately after sources switch on.   We must integrate these equations to calculate the temperature and ionization fluctuations at a given redshift and for a given value of $k$.

These equations illuminate the effect of heating.  First, consider $g_T$, which we can easily relate to the adiabatic index of the gas $\gamma_a$ by $g_T=\gamma_a-1$, giving it a simple physical interpretation.  Adiabatic expansion and cooling tends to drive $g_T\rightarrow2/3$ (corresponding to $\gamma_a=5/3$, appropriate for a monoatomic ideal gas), but when Compton heating is effective at high $z$, it deposits an equal amount of heat per particle, driving the gas towards isothermality ($g_T\rightarrow0$).  At low $z$, where X-ray heating of the gas becomes significant, the temperature fluctuations are dominated by spatial variation in the heating rate ($g_T\rightarrow W_X$).  This embodies the higher temperatures closer to clustered sources of X-ray emission.  If the heating rate is uniform $W_X(k)\approx0$, then the spatially constant input of energy drives the gas towards isothermality $g_T\rightarrow0$.   

The behaviour of $g_e$ is similarly straightforward to interpret.  At high redshift, when the IGM is dense and largely neutral, the ionization fraction is dominated by the recombination rate, pushing $g_x\rightarrow-1$, because denser regions recombine more quickly.  As the density decreases and recombination becomes ineffective, the first term of equation \eqref{evolve_ge2} 
slowly pushes $g_x\rightarrow0$.  Again, once ionization becomes important, the ionization fraction is pushed towards tracking spatial variation in the ionization rate ($g_x\rightarrow W_X$).
Note that, because the ionization fraction in the bulk remains less than a few percent, fluctuations in the neutral fraction remain negligibly small at all times. 

The scale dependence of $g_T$ is illustrated in Figure \ref{fig:wkcomp}.  $g_T$ tries to track the heating fluctuations $W_X(k)$ (as in the $z=15$ curve), but two factors prevent this.  First, until heating is significant, the effect of adiabatic expansion tends to smooth out variations in $g_T$.  Second, $g_T$ responds to the integrated history of the heating fluctuations, so that it tends to lag $W_X$ somewhat.  When the bulk of star formation has occurred recently, as when the star formation rate is increasing with time, then there is little lag between $g_T$ and $W_X$.  In contrast, when the star formation rate has reached a plateau or is decreasing the bulk of the X-ray flux originates from noticeably higher $z$ and so $g_T$ tends to track the value of $W_X$ at this higher redshift. On small scales, the heating fluctuations are negligible and $g_T$ returns to the value of the (scale independent) uniform heating case.

\section{Temperature Fluctuations} 
\label{sec:tkfluctuate}

\begin{figure}
\begin{center}
\includegraphics[scale=0.4]{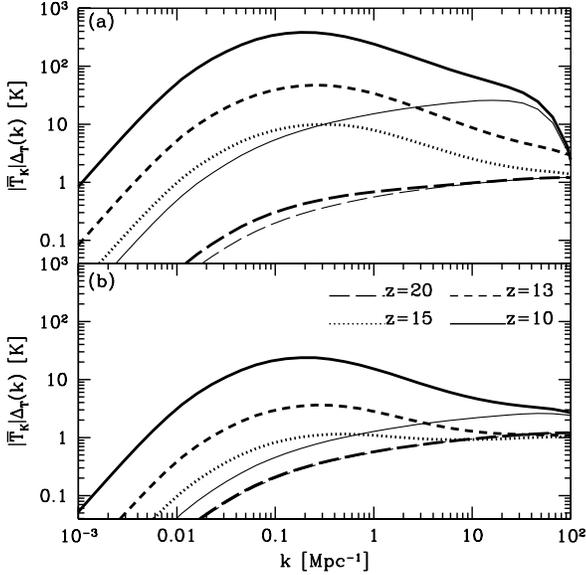}
\caption{Fluctuations in $T_K$. In each panel we plot $\bar{T}_K\Delta_T(k)$ for the case of  inhomogeneous X-ray heating (thick curves) at $z=20$ (long dashed curve), $z=15$ (dotted curve), $z=13$ (short dashed curve), and $z=10$ (solid curve). For comparison, we plot the case of uniform heating at $z=10$ (thin solid curve) and $z=20$ (thin long dashed curve). {\em (a): } Model A. {\em (b): } Model B.}
\label{fig:tk_full}
\end{center}
\end{figure}
Before calculating the 21 cm signal, let us first examine the gas temperature fluctuations themselves.  Figure \ref{fig:tk_full} shows  the power spectrum of temperature fluctuations $P_T(k)$ for models A and B respectively.  We see that in both cases the fluctuations are small until $z<20$.  
At lower redshifts and on larger scales ($k\approx0.1\iMpc$), the heating fluctuations source a significant (factor of $\approx 50$) enhancement over the uniform heating case.  This is to be expected.  Uniform heating of the gas tends to erase temperature fluctuations, while inhomogeneous heating causes them to grow.  Thus we observe a huge increase in power.  The fluctuation amplitude in Model B is generally smaller than in Model A as a consequence of the reduced heating from the decreased SFR in Model B.  In both cases, the temperature fluctuations remain small, $\delta_T<1$ (compare with Figure \ref{fig:global1}), justifying our linear approximations.

\begin{figure}
\begin{center}
\includegraphics[scale=0.4]{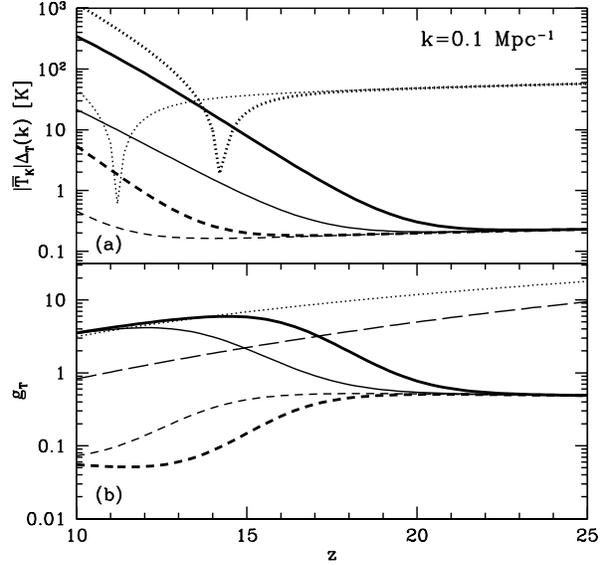}
\caption{Evolution of the fluctuations in $T_K$ with redshift for Model A (thick curves) and B (thin curves). {\em (a): }We plot the amplitude of fluctuations in $T_K$ given by $\bar{T}_K\Delta_T(k)$ at $k=0.1\iMpc$ in the case of uniform heating (dashed curves) and when fluctuations in the heating rate are considered (solid curves). For comparison, we plot $|\bar{T}_k-T_\gamma|$ (dotted curves).  Only in a small region of width $\Delta z\approx1$ around $\bar{T}_K=T_\gamma$ do the fluctuations exceed this threshold.  {\em (b): }  Evolution of $g_T$.  We plot $g_T$ at $k=0.1\iMpc$ for the uniform (short dashed curves) and fluctuating cases (solid curves).  We also plot $W_X$ (dotted curve) and, for comparison, $W_{\alpha,\star}(k)$ (long dashed curve).  Notice how $g_T$ rises to track $W_X$ once heating becomes effective. }
\label{fig:tkfluctuateZ}
\end{center}
\end{figure}
Figure \ref{fig:tkfluctuateZ} illustrates the redshift evolution of the temperature fluctuations.  We choose to follow a single wavenumber $k=0.1\iMpc$, which is both within those scales accessible to future experiments and demonstrative of the effect.  If the gas is heated uniformly (dashed curves), then $g_T$ rapidly becomes negligible once heating becomes effective.  By depositing the same amount of energy per particle the gas is driven towards isothermality. When heating fluctuations are taken into account $g_T$ may grow or decrease depending on scale.  We observe that, for the scale chosen here, the amplitude of the temperature fluctuations grows steadily with time, but $g_T$ decreases.  This is a consequence of the sources becoming less biased with time so that $W_X(z)$ decreases with $z$.  On very small scales, where $W_X(k)$ is negligible, $g_T$ will trace the uniform heating curve.

Recall that whether we observe the 21 cm line in emission or absorption depends on the sign of $T_S-T_\gamma$.  Assuming that $T_S\approx T_K$, when $T_K<T_\gamma$, hotter regions have a spin temperature closer to the CMB temperature and so appear more faintly in absorption.  As heating continues, it is these regions that are first seen in emission, when their temperature exceeds $T_\gamma$.  Once $T_K>T_\gamma$, these hotter regions produce the largest emission signal. 

We see from Figure \ref{fig:tkfluctuateZ} that for a short window around $z_h$ (where $\bar{T}_K=T_\gamma$) temperature fluctuations may raise $T_K$ above $T_\gamma$ in these hot regions, even when $\bar{T}_K$ is less than $T_\gamma$.  We interpret this to mean that within this window the 21 cm signal will be a confusing mix of emission, from hotter regions, and absorption, from cooler regions.  In the case of uniform heating this window is very narrow, but when fluctuations are included it extends to a significant ($\Delta z\approx1$) width.  This indicates that the transition from absorption to emission will not be abrupt, but extended.  

\section{21 cm Power Spectrum} 
\label{sec:tbfluctuate}
\subsection{Redshift evolution}
\label{ssec:evolution}
Finally, we write the full 21 cm power spectrum as
\begin{equation}
P_{T_b}(k,\mu)=\bar{T}_b^2(\beta'+\mu^2)^2P_{\delta\delta}(k),
\end{equation}
where
\begin{equation}
\beta'=\beta-\beta_x \bar{x}_e g_e/(1+\bar{x}_e)+\beta_T g_T+\beta_\alpha W_\alpha.
\end{equation}
Within our model we may neglect the term corresponding to the neutral fraction, as the free electron fraction in the IGM remains small at all times.  We now consider how the 21 cm power spectrum evolves with redshift.

\begin{figure}
\begin{center}
\includegraphics[scale=0.4]{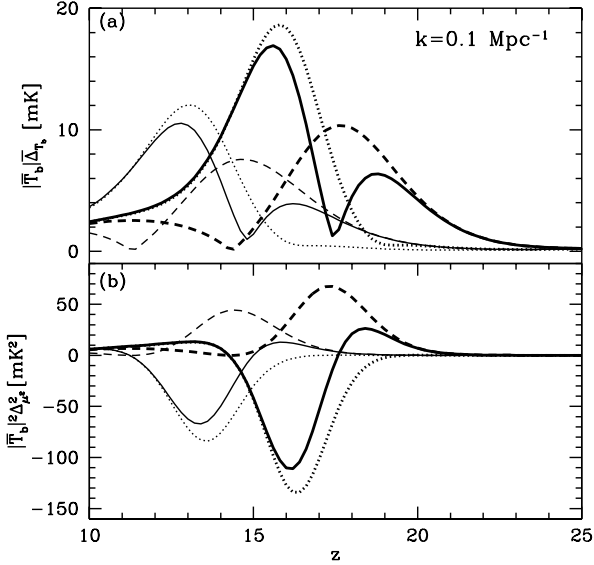}
\caption{Evolution of brightness temperature fluctuations for Model A (thick curves) and B (thin curves). {\em (a): }  We plot $|\bar{T}_b|\bar{\Delta}_{T_b}(k)$ at $k=0.1\iMpc$ including the effects of heating fluctuations (dotted curves), \lya fluctuations (dashed curves), and both heating and \lya fluctuations (solid curves).  {\em (b): }We plot $|\bar{T}_b|^2\Delta_{\mu^2}^2(k)$ with the same line conventions.}
\label{fig:tbfluctuateZ}
\end{center}
\end{figure}
Figure \ref{fig:tbfluctuateZ} shows the evolution of the brightness temperature fluctuations at a single scale $k=0.1\iMpc$ with redshift.  First, note that in the bottom panel $\Delta^2_{\mu^2}$ changes sign when we include temperature fluctuations (note that $\Delta_{\mu^2}^2$ is not an auto-correlation and so is free to have a negative sign).  Physically, this occurs because when $T_K<T_\gamma$ there is an anti-correlation between $T_b$ and $T_K$, i.e. increasing $T_K$ decreases $T_b$.  Observing $P_{\mu^2}<0$ is a clear sign that $T_K<T_\gamma$.  Mathematically, this can be seen because $\beta_T$ is the only one of the fluctuation coefficients that can become negative. Of course, if $P_{\delta T}$ or other cross-correlations become negative we can also get $P_{\mu^2}<0$, but this should not be the case for radiative heating or \lya coupling, as we expect emitting sources to be most common in overdense regions.  Only in the case of $P_{x\delta}$ might we expect a negative cross-correlation, as increasing the UV radiation is likely to decrease the neutral fraction.  In the high redshift regime, before significant ionization has occurred, this term is negligible.

Adding the \lya fluctuations, we see a clear double peaked temporal structure in the evolution of $\bar{\Delta}_{T_b}$, which is dominated by \lya fluctuations at high $z$ and temperature fluctuations at lower $z$ (were we to include the effects of ionization fluctuations, there would be a third peak at still lower redshift). We note that there is considerable overlap between the two signals, which will complicate extracting astrophysical information.  The situation is similar in Model B, although here the relevant signal is compressed into a narrower redshift window. We note that the amplitude of fluctuations induced by the gas temperature is significantly larger than those from the \lya signal and present at lower redshifts.  Both of these features make the temperature fluctuation signal a plausible target for future observations.

To illustrate the scale dependence of this signal, we examine a series of redshift slices.  We will make plots for model A.  Although the same evolution applies for model B, the events are shifted to lower redshift $\Delta z\approx3$ and the transitions are somewhat compressed in redshift.  We begin by examining the high redshift regime, where \lya fluctuations dominate the 21 cm signal, but temperature fluctuations become important as we move to lower redshift.

\begin{figure}
\begin{center}
\includegraphics[scale=0.4]{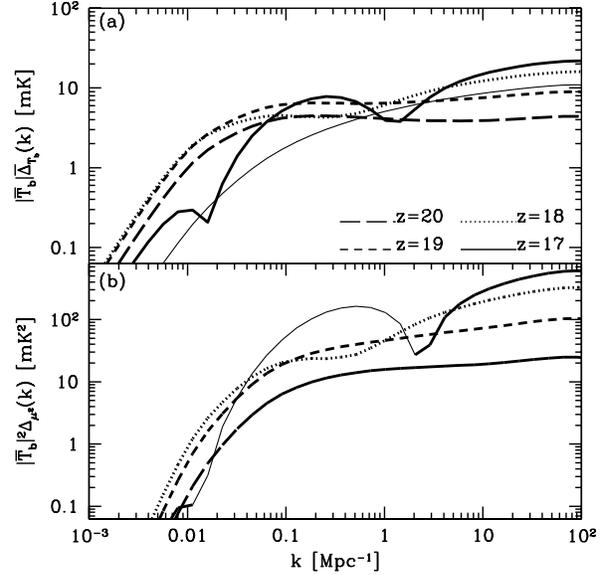}
\caption{Full $T_b$ power spectra for Model A.  We plot the power spectra at $z=20$ (long dashed curve), $z=19$ (short dashed curve), $z=18$ (dotted curve), and $z=17$ (solid curve). {\em (a): }$|\bar{T}_b|\bar{\Delta}_{T_b}(k)$.  We plot $|\bar{T}_b|\Delta_{\delta\delta}$ (thin solid curve) at $z=19$ for comparison.  {\em (b): }$|\bar{T}_b|^2\Delta^2_{\mu^2}(k)$.  The sign of $\Delta^2_{\mu^2}(k)$ is indicated as positive (thick curves) or negative (thin curves).}
\label{fig:tb_full_slice1}
\end{center}
\end{figure}
Figure \ref{fig:tb_full_slice1} shows redshift slices from $z=17-20$.  We can see from Figure \ref{fig:tbfluctuateZ} that \lya fluctuations dominate the signal for $z\gtrsim18$.  The $z=20$ and $z=19$ curves show the expected excess of power on large scales for \lya fluctuations from the first sources (see \citealt{bl2005detect} for a full analysis of this signal).  At $z=18$, we begin to see the effects of the temperature fluctuations through the dip in power between $k=0.1$ and $1\iMpc$.  This dip occurs because $\beta_T<0$, contrasting with the other $\beta_i$, which are positive.  Physically, in this regime $T_K<T_\gamma$ and regions that are hotter have a smaller brightness temperature.  In our model, denser regions are more strongly coupled, which increases $T_b$, but are also hotter, which tends to decrease $T_b$.  These two effects compete with one another and produce the dip.

At $z=17$, temperature fluctuations grow large enough to drive $\beta'$ negative over a range of scales, where they outweigh the \lya fluctuations.  This leads to a sign change in $\Delta_{\mu^2}^2$, but also imprints a distinctive trough-peak-trough structure in $\bar{\Delta}_{T_b}$.  Here \lya fluctuations dominate on the largest scales, temperature fluctuations on intermediate scales, and density fluctuations on small scales.  For this to occur, we require that $W_\alpha>g_T$ on large scales, which can only occur if $W_\alpha$ and $g_T$ show different scale dependence.  This always occurs at some redshift in our model, as both $W_\alpha$ and $W_X$ tend towards the same value on large scales, but $g_T$ lags behind (and so is smaller than) $W_X$ on those scales.

\begin{figure}
\begin{center}
\includegraphics[scale=0.4]{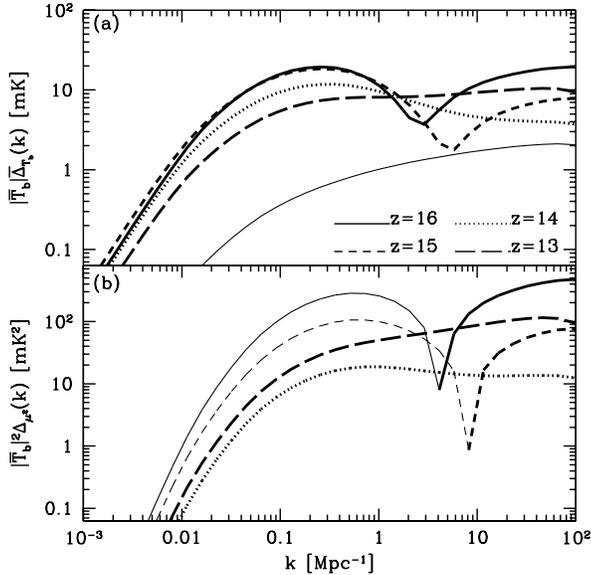}
\caption{Full $T_b$ power spectra for Model A.  We plot the power spectra at $z=16$ (solid curve), $z=15$ (short dashed curve), $z=14$ (dotted curve), and $z=13$ (long dashed curve). Note that the $z=13$ curve would be seen in emission, and the $z=14$ curve in a mixture of emission and absorption.  The other curves would be seen in absorption against the CMB.  {\em (a): }$|\bar{T}_b|\bar{\Delta}_{T_b}(k)$.  We plot $|\bar{T}_b|\Delta_{\delta\delta}$ (thin solid curve) at $z=14$ for comparison.  {\em (b): }$|\bar{T}_b|^2\Delta^2_{\mu^2}(k)$.  The sign of $\Delta^2_{\mu^2}(k)$ is indicated as positive (thick curves) or negative (thin curves). }
\label{fig:tb_full_slice2}
\end{center}
\end{figure}
From Figure \ref{fig:tbfluctuateZ}, we see that $T_K$ fluctuations dominate at $z\lesssim17$ and that \lya fluctuations become negligible for $z\lesssim15$.  In Figure \ref{fig:tb_full_slice2} we plot redshift slices in the range $z=13-16$.  At $z=16$ and $z=15$, we see a sign change in $\Delta_{\mu^2}$, which is a distinctive signature of the temperature fluctuations when $T_K<T_\gamma$.  This is seen in $\bar{\Delta}_{T_b}$ as a peak on large scales, followed by a trough at smaller scales.  The position of the peak depends upon the shape of $g_T$ and thus the X-ray source spectrum.  We will consider this in more detail in the next section.  

Notice that the heating transition occurs very close to $z=14$, so that the 21 cm signal at this redshift would likely be seen in a mixture of absorption and emission.  In addition, this curve is dominated by gas temperature fluctuations.  We see this in Figure \ref{fig:tb_full_slice2} where the contribution from density fluctuations at $z=14$ (thin solid curve) is at least a factor of two smaller than $\bar{\Delta}_{T_b}$ on all scales.  Recall from Figure \ref{fig:global2} that when $\bar{T}_b\approx0$ only the combination $|\bar{T}_b|\beta_T$ is significant.  

The position of the sign change moves to smaller scales as the gas is heated and the temperature fluctuations become larger.  Eventually, the IGM heats to $\bar{T}_K>T_\gamma$, hotter regions have a higher brightness temperature than average and $\beta_T>0$.  Once this occurs the trough disappears entirely and the peak on large scales is no longer quite so distinctive (see $z=14$ curve).  The continued IGM heating drives $\beta_T\rightarrow0$ and diminishes the effect of the temperature fluctuations.  By $z=13$ there is no longer a clear peak in either $\Delta^2_{\mu^2}$ or $\bar{\Delta}_{T_b}$, although there is still considerable excess power on large scales.  By $z=10$, $T_K\gg T_\gamma$ and temperature fluctuations no longer impact the 21 cm signal significantly. 

Once the ionization fraction becomes large ($x_i\gtrsim 0.1$), the 21 cm signal becomes dominated by the imprint of HII regions \citep{zfh2004freq,fzh2004}.  This eventually produces a distinct knee in the 21 cm power spectrum resulting from the characteristic size of the bubbles.  We note that our models have $x_i\lesssim0.1$ at $z\gtrsim12$, so that we do not expect ionization fluctuations to significantly affect the results we have outlined for Model A.  In the case of model B, temperature fluctuations remain significant to lower redshift where they may interfere with attempts to measure the power spectrum of ionization fluctuations. The reverse is also true.

\subsection{Spectral dependence}
\label{ssec:xrayspectrum}

We next imagine using the temperature fluctuations to constrain the X-ray source spectra.  This should affect the temperature fluctuations on intermediate scales, where heating fluctuations dominate.  Increasing the hardness of the spectrum increases the fraction of more energetic photons, which have longer mean free paths.  This should further smooth the temperature fluctuations and suppress power on small scales. 

\begin{figure}
\begin{center}
\includegraphics[scale=0.4]{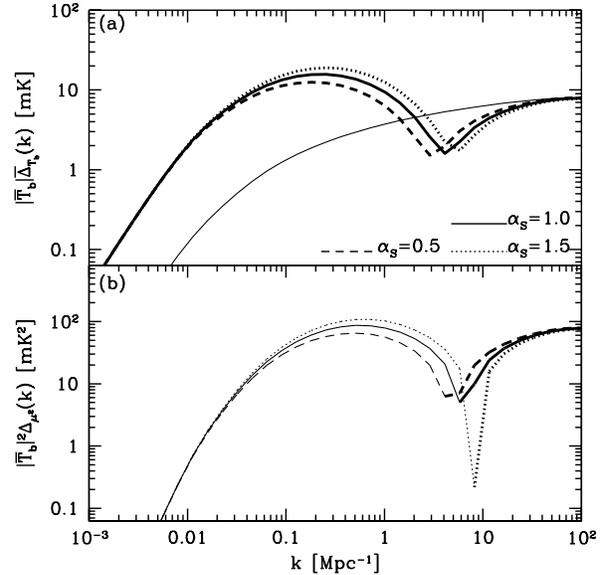}
\caption{Effect of X-ray spectra on 21 cm power spectra.  We show results at $z=15$ for model A and take $\alpha=1.5$ (dotted curve), $\alpha=1.0$ (solid curve), and $\alpha=0.5$ (dashed curve). {\em (a): }$|\bar{T}_b|\bar{\Delta}_{T_b}(k)$.  We illustrate the uniform heating case by the thin solid curve. {\em (b): }$|\bar{T}_b|^2\Delta^2_{\mu^2}(k)$. }
\label{fig:tb_specX_A}
\end{center}
\end{figure}
Figure \ref{fig:tb_specX_A} shows the power spectra at $z=15$ (chosen to maximise the distinctive features of the temperature fluctuations) for source spectra $\alpha_S=1.5$ (mini-quasars), $\alpha_S=1.0$ (SNR), and $\alpha_S=0.5$.  We see that the spectra alter the most on scales $k\approx0.1-10\iMpc$.  The two main signatures are the change in amplitude and shift in the position of the trough.  Both of these occur because increasing the slope of the spectrum, with fixed total luminosity, increases the number of soft X-rays and so increases the heating in smaller scales.  The trough (or sign change in $P_{\mu^2}$) shifts by $\Delta k\sim2\iMpc$ for $\Delta\alpha=0.5$, an effect that might be observable were it not located on small scales $k\approx5\iMpc$.  The amplitude change at the peak is more observable but is also degenerate with modifications in the thermal history, making this a very challenging measurement to perform in practice.

Referring back to our discussion of the time evolution of the signal, we see that this sort of variation is similar to the effect of changing the thermal history.  However, the exact shape of the spectrum is determined by the form of $g_T$, and hence $W_X$.  These do encode distinct information about the source spectrum.  Consequently, precision measurements of the 21 cm power spectrum at high $z$ could constrain the X-ray source spectrum.

We can also seek to constrain the X-ray spectrum by looking at the regime where fluctuations in the \lya flux dominate the 21 cm signal.  The inclusion of \lya photons generated by X-ray excitation of HI (in addition to those redshifting into the Lyman resonances) modifies the shape of the power spectrum significantly.  This is easy to see by referring back to Figure \ref{fig:wkcomp}.  There we plotted $W_{\alpha,\star}(k)$, for the case of stellar emission, and $W_X(k)$, which determines the fluctuations in the X-ray flux.  If we allow both stars and \lya photons produced from X-rays to contribute to the \lya flux, then the resulting spectrum of fluctuations is determined by a weighted combination of these $W_\alpha(k)$ and $W_X(k)$.  In our model, as in that of \citet{chen2006}, the \lya flux is dominated on small scales by the X-ray contribution and on large scales by the stellar contribution.  Thus the resulting weighting function most closely resembles $W_X(k)$ with significant power on small scales.  

\begin{figure}
\begin{center}
\includegraphics[scale=0.4]{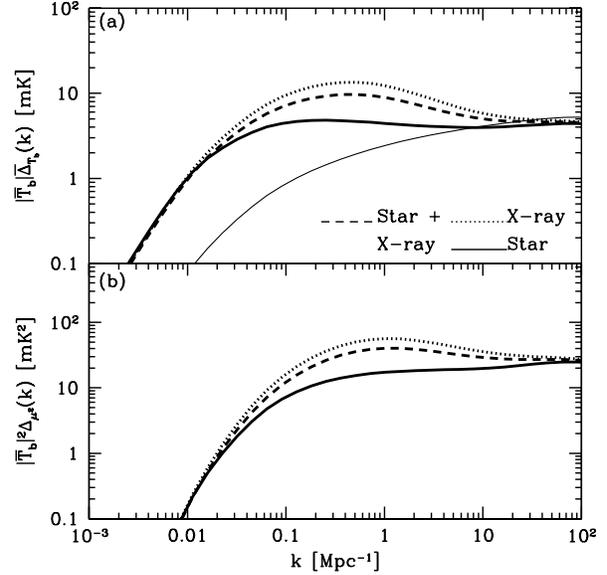}
\caption{{\em (a): } $|\bar{T}_b|\bar{\Delta}_{T_b}(k)$. We consider the following sources of \lya emssion: stellar only (solid curve), X-ray excitation only (dotted curve), stellar+X-ray excitation (dashed curve).  All curves are calculated at $z=20$ and have been normalised to the stellar only case, to compensate for different mean values of $x_\alpha$.  We assume X-ray emission from starburst galaxies. Also plotted is $|\bar{T}_b|\Delta_{\delta\delta}$ (thin solid curve). {\em (b): } $|\bar{T}_b|^2\Delta^2_{\mu^2}(k)$. Same line conventions as in {\em (a)}.}
\label{fig:tb_specL_A}
\end{center}
\end{figure}
Figure \ref{fig:tb_specL_A} shows the effect on the power spectrum at $z=20$, when temperature fluctuations are negligible, of including the different contributions to the \lya flux.  On intermediate scales ($k\approx1\iMpc$) there is clearly significantly more power when X-ray excitation dominates \lya production compared to stellar production.  As noted in \citet{chuzhoy2006}, this provides a means for distinguishing between the major source of \lya photons during the time of the first sources.  We note that the shape of the spectrum is somewhat sensitive to the spectral index of the X-ray sources - with the variation being similar to between the stellar + X-ray and X-ray only curves.  Thus isolating the 21 cm fluctuations from the \lya flux variations could also constrain the X-ray spectrum of the first sources.

\subsection{Effects of X-ray background}
\label{ssec:xray}
\begin{figure}
\begin{center}
\includegraphics[scale=0.4]{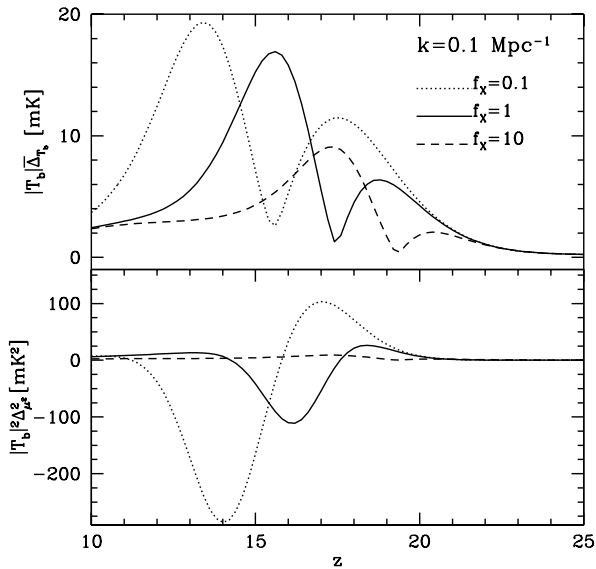}
\caption{{\em (a): }Redshift evolution of $|\bar{T}_b|\bar{\Delta}_{T_b}(k)$ at $k=0.1\iMpc$ for model A, but with $f_X=0.1$ (dotted curve), 1.0 (solid curve), and 10 (dashed curve)  {\em (b): }Redshift evolution of $|\bar{T}_b|\Delta_{\mu^2}(k)$.  Same line conventions as in {\em (a)}. }
\label{fig:fx}
\end{center}
\end{figure}

We now explore the effect of modifying the X-ray luminosity of our sources.  We have so far taken $f_X=1$ in our analysis, but constraints on the high redshift X-ray background are weak giving us significant freedom to vary $f_X$, which parametrizes the source luminosity. As an example, for our model A, values of $f_X\lesssim 10^{3}$ are easily possible without X-ray or collisional ionization of the IGM violating WMAP3 constraints on $\tau$ at the 2-sigma level.  In Figure \ref{fig:fx}, we show the time evolution of the 21 cm fluctuations for model A, taking $f_X=0.1$, 1, and 10.  This serves to illustrate the effect of late or early X-ray heating and illustrates the range of uncertainty in making predictions. 

Earlier heating (dashed curve) causes the temperature fluctuations to become important at higher redshift, cutting into the region of \lya fluctuation.  This will make the 21 cm signal more complicated as temperature and \lya fluctuations contribute over a similar range of redshifts.  However, early heating also means that temperature fluctuations become unimportant for the 21 cm signal at late times improving the prospects for extracting cosmology from the 21 cm signal \citep{mcquinn2005,santos2005}.  In contrast, late heating (dotted curve) allows a clearer separation between temperature and \lya fluctuations, but means temperature fluctuations are likely to be important during the beginning of reionization.  This will complicate the extraction of information about HII regions as reionization gets underway.  

Clearly there is considerable uncertainty as to the behaviour of the 21 cm signal at high redshifts due to our poor understanding of the source populations.  Viewed another way, measurement of the evolution of the 21 cm signal could provide useful constraints on the X-ray background at high redshift.  This is important as efforts to observe the diffuse X-ray background are complicated by technical issues of calibration.  We also note that for weaker X-ray heating other sources of heating, especially shock heating, may become important.

Finally we remind the reader that our model is applicable in the IGM outside of ionized HII regions.  If heating occurs late, so that temperature fluctuations are important as HII regions become large, then it will be important to extend this model if accurate predictions of the 21 cm signal during reionization are to be made.  It will also be important to include these temperature fluctuations into simulated predictions of the 21 cm signal.  

\section{Observational Prospects} 
\label{sec:observe}

We now turn to the important question of observing the features outlined above.  The first generation of 21 cm experiments (PAST, LOFAR, MWA) will be optimised to look for the signature of HII regions at redshifts $z\lesssim12$.  Their sensitivity decreases rapidly at redshifts  $z\gtrsim10$ (\citealt{mcquinn2005}; \citealt*{bowman2006}) and so they are unlikely to be able to detect the effects of inhomogeneous heating.  The proposed successor to these instruments, the SKA, is still under design, but its fiducial specifications should allow the $z>12$ regime to be probed.  In this section, we will consider using an SKA type experiment to observe 21 cm fluctuations at $z=13$ and $z=15$ and calculate the achievable precision.

Before this, we must make the necessary caveats concerning foregrounds.  Foregrounds for 21 cm observations include terrestrial radio interference (RFI), galactic synchrotron emission, radio recombination lines, and many others (\citealt*{ohmack2003,dimatteo2004}; \S9 of \citealt{fob}).  Typical foregrounds produce system temperatures $T_{\rm{sys}}\gtrsim1000 \,\rm{K}$, compared to a signal measured in $\rm{mK}$.  These foregrounds increase rapidly as we move to lower frequency, making their removal an even greater concern for high redshift observations than low ones.  Although techniques for foreground removal are well grounded, their effectiveness has yet to be tested.  In the analysis that follows, we assume that foreground removal can be effected by exploiting the smoothness of foregrounds in frequency space (\citealt*{zfh2004freq,santos2005,morales_hewitt2004,morales2005}; \citealt{mcquinn2005,wang2006}).  

\begin{figure}
\begin{center}
\includegraphics[scale=0.4]{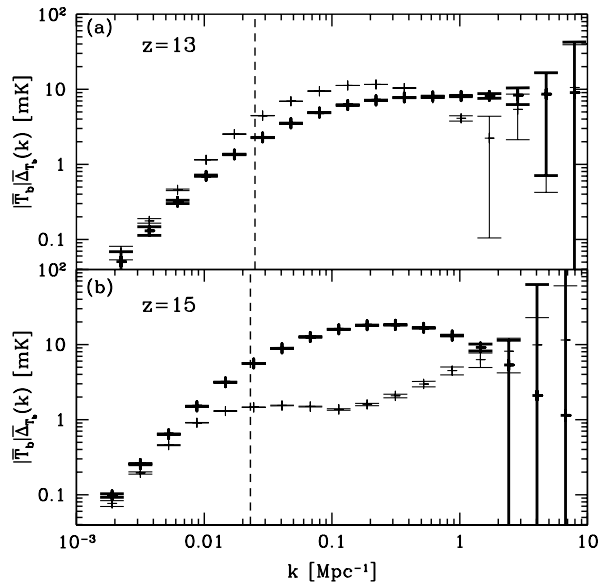}
\caption{Predicted 1-$\sigma$ errors on $|\bar{T}_b|\bar{\Delta}_{T_b}(k)$ for an SKA like instrument (see text for details).  We compare spectra for model A (thick curves) and model B (thin curves).  Modes with $k\lesssim k_{\rm{foreground}}$ (shown by a vertical dashed line) will probably be lost during foreground cleaning. {\em (a): } $z=13$. {\em (b): } $z=15$.  }
\label{fig:errorplot}
\end{center}
\end{figure}
Figure \ref{fig:errorplot} shows predicted $1-\sigma$ error bars on $\bar{\Delta}_{T_b}(k)$ at $z=13$ and $z=15$ for model A and B.  We assume an SKA like instrument with a total effective area $A_{\rm{tot}}=1\,\rm{km}^2$ distributed over $5000$ antennae in a 5 km core, bandwidth $B=12 \,\rm{MHz}$, minimum baseline $D_{\rm{min}}=10\,\rm{m}$, and an integration time $t_{\rm{int}}=1000\,\rm{hr}$.  We set $T_{\rm{sys}}=1000\,\rm{K}$ and $1400\,\rm{K}$ at these two redshifts and use bins of width $\Delta k=k/2$.  We assume that foregrounds can be removed exactly, but that this also removes cosmological information on scales exceeding the bandwidth of the observations, so that modes with $k\le k_{\rm{foreground}}\approx0.025\iMpc$ (indicated by vertical dashed lines) are lost \citep{mcquinn2005}.

With these caveats, observations could measure $\bar{\Delta}_{T_b}(k)$ accurately over the range $k\approx0.025-2\iMpc$.  The precision is more than adequate to distinguish between model A and B.  Detecting the characteristic peak-trough signature of $\delta_T$ is difficult, as the trough typically occurs on small scales where the uncertainty is large.  However, it should be possible to detect the peak and the beginning of the decline.  We note that detection of the trough is necessary to unambiguously determine $\beta_T<0$ and so show that $T_S<T_\gamma$.  Without this it is not simple to distinguish between the two cases exemplified by the $z=14$ curve, which has no trough, and the $z=15$ curve, which does, in Figure \ref{fig:tb_full_slice2}.  No similar confusion occurs when the reduction in power caused by $\beta_T<0$ is obvious, as in the $z=15$ model B case.  

From the point of view of constraining the spectra of X-ray sources, the precision is adequate for distinguishing between the different curves of Figure \ref{fig:tb_specL_A}.  Whether the effect of the spectrum can be separated out from different thermal histories is an open question, which deserves future study.

Throughout this work, we have ignored the effect of the HII regions on the 21 cm power spectrum.  While this is reasonable at high redshifts, this approximation will begin to break down as the filling fraction of ionized regions increases.  The bubble model of \citet{fzh2004} predicts that these bubbles remain at sub-Mpc sizes while $x_i\lesssim0.1$.  Consequently, we naively expect contamination of the signal by these bubbles to be confined to small scale modes with $k\gtrsim 1 \iMpc$ that will be very difficult to detect.  Exploring the detailed interaction between  temperature and neutral fraction fluctuations is beyond the scope of this paper, but may be important for detailed predictions of the 21 cm signal at the beginning of reionization.

\citet{santos2006} have considered the extraction of astrophysical and cosmological parameters from 21 cm observations in the period of the first sources.  They assumed that gas temperature fluctuations showed no scale dependence $g_T(k,z)=g_T(z)$ and argued that extracting astrophysical information using an SKA like instrument is difficult but feasible.  We expect the scale dependent temperature fluctuations that we have investigated to both help and hinder parameter estimation.  Figure \ref{fig:errorplot} shows that it should be possible to resolve individual features imprinted in the power spectrum by temperature fluctuations.  These features provide additional leverage in extracting astrophysical parameters.  However, the shape of the power spectrum evolves rapidly in our model, making binning of different redshift data more difficult.

\section{Conclusions} 
\label{sec:conclusion}
X-ray production by an early generation of stellar remnants is widely regarded as the most likely candidate for heating the IGM above the CMB temperature from its cool adiabatic level.  This heating has often been treated as uniform, as the mean free path of hard X-rays in the early Universe is comparable to the Hubble scale.  We have relaxed this assumption and, by expanding on the formalism of \citet{bl2005detect}, calculated the temperature fluctuations that arise from the inhomogeneous heating.  The spectrum of fluctuations in $T_K$ is significantly larger than that predicted from uniform heating, peaking on scales $k\approx0.1\iMpc$.  This allowed us to examine the redshift range about $z_h$, where $T_K=T_\gamma$, and show that there is a window of width $\Delta z\approx1$ in which the IGM will contain pockets of gas both hotter and colder than the CMB.  This has implications for the 21 cm signal, which will be seen in a mixture of absorption and emission within this window.

The best hope for observing the temperature evolution before reionization is through 21 cm observations of neutral hydrogen.  Systematic effects arising from foregrounds are likely to prevent interferometers from measuring $\bar{T}_b$ directly \citep{fob}, although several alternative methods for obtaining $\bar{T}_b$ have been proposed \citep{bl2005sep,cooray2006}.  Thus careful analysis of brightness fluctuations will be required to extract astrophysical information.  Fluctuations in $T_K$ lead to fluctuations in $T_b$, which contain information about the thermal history and the nature of the heating sources.  We have calculated the 21 cm power spectrum arising from inhomogeneous X-ray heating and shown that it has considerable structure.  In the regime where gas temperature and \lya flux fluctuations compete, we expect a trough-peak-trough structure in $\bar{\Delta}_{T_b}(k)$.  Once $T_K$ fluctuations dominate, but while $T_K<T_\gamma$, we see a peak-trough structure.  As the gas heats, this structure is lost as the trough moves to unobservable small scales while the peak decreases and finally vanishes once $T_K\gg T_\gamma$.  Extracting astrophysical information cleanly will be challenging, but the information is there.  

It is important to notice that the difference between uniform and inhomogeneous heating is large.  Observations with the SKA should be able to distinguish these two cases and indicate whether X-ray heating is important.  If it is possible to perform an angular separation of $P_{T_b}$, then observing $P_{\mu^2}<0$ is a clear indicator that $T_K<T_\gamma$.  Ideally, one would extract the quantity $\beta_T$, but this requires fitting of other parameters and so is a less direct (but more conclusive) observational feature.

Additionally, the spectra of the X-ray sources imprint information on the $T_K$ fluctuations.  This may be observed in the 21 cm power spectra, where it shifts the critical scale at which $P_{\mu^2}$ changes sign, or during the regime in which \lya fluctuations dominate, where it modifies the shape of the power spectrum.  The temperature fluctuations that we have calculated lead to a 21 cm signal that extends down to relatively low redshifts.  This opens an opportunity for future 21 cm radio arrays to probe the thermal history prior to reionization.  Including temperature fluctuations makes the 21 cm signal significantly more complex, adding information, but further raises the question of how best to separate out that information.

In this paper, we have ignored the contribution from Poisson fluctuations in the source distribution \citep{bl2005detect}.  While calculating it requires only a straightforward extension of the \citet{bl2005detect} formalism, performing the time integrals necessary to convert heating fluctuations into temperature fluctuations is non-trivial.  We have estimated the amplitude of these Poisson temperature fluctuations and find them (in our models) to be subdominant at all redshifts.  This is largely because there are many more sources at the lower redshifts where temperature fluctuations are important.  In theory, high-precision 21 cm observations can separate these Poisson fluctuations from fluctuations correlated with the density field.  The Poisson contribution could then be used to probe the distribution of sources, for example, by distinguishing between highly biased mini-quasars and less biased star-burst galaxies, producing the same global X-ray luminosity.

In our analysis we have taken $f_X=1$, corresponding to normalising the X-ray luminosity per unit star formation to that observed in the local universe.  In truth, this assumption is highly speculative and the value for $f_X$ is extremely uncertain. We have investigated the effects of changing $f_X$ and find that it alters the details of the thermal evolution significantly.  Taking $f_X=0.1$, for example, shifts the point where 21 cm brightness fluctuations change from being dominated by \lya fluctuations to gas temperature fluctuations from $z\approx17$ to $z\approx15$.  Setting $f_X=10$ increases the redshift of this transition to $z\approx19$.  For values of $f_X\lesssim0.1$, we find a clear separation between 21 cm brightness fluctuations sourced by gas temperature and \lya fluctuations.  Increasing $f_X$ also increases the redshift at which $\bar{T}_K\gg T_\gamma$, so that gas temperature fluctuations become irrelevant for the 21 cm signal.  Additionally, small values for $f_X$ will increase the contribution of other heating mechanisms such as shock heating.  All of this suggests that measuring the time evolution of the 21 cm signal (as in Figure \ref{fig:fx} for example) would enable $f_X$ to be constrained.  Unfortunately, until these observations are made it is difficult to predict the thermal history before reionization with any certainty.

We have shown that the 21 cm signal at high $z$ will contain significantly more structure than has previously been considered.  Temperature fluctuations produce an interesting interplay with other sources of 21 cm anisotropy as $\beta_T<0$ when $T_K<T_\gamma$.  Furthermore, for reasonable heating scenarios, the effect of temperature fluctuations persist well into the regime that will be probed by second generation low-frequency arrays, such as the SKA.  
Thus, prospects for probing the thermal history before reionization via observations of the redshifted 21 cm line seem promising.

This work was supported at Caltech in part by DoE DE-FG03-92-ER40701.  JRP would like to thanks Miguel Morales for useful discussions.  SRF thanks the Tapir group at Caltech for hospitality while this work was completed.  We would also like to thank the anonymous referee for many useful comments that have helped improve the clarity of the paper.



\end{document}